# Spin-Phonon Coupling and High-Temperature Phase Transition in Multiferroic Material YMnO$_3$

Mayanak Kumar Gupta[1], Ranjan Mittal[1], Mohamed Zbiri[2], Neetika Sharma[1], Stephane Rols[2], Helmut Schober[2] and Samrath Lal Chaplot[1]

[1]*Solid State Physics Division, Bhabha Atomic Research Centre, Mumbai 400085, India*
[2]*Institut Laue-Langevin, 71 Avenue des Martyrs, CS 20156, 38042 Grenoble Cedex 9, France*

We have carried out temperature-dependent inelastic neutron scattering measurements of YMnO$_3$ over the temperature range 50 - 1303 K, covering both the antiferromagnetic to paramagnetic transition (70 K), as well as the ferroelectric to paraelectric transition (1258 K). Measurements are accompanied by first principles calculations of phonon spectra for the sake of interpretation and analysis of the measured phonon spectra in the room temperature ferroelectric (P6$_3$cm) and high temperature paraelectric (P6$_3$/mmc) hexagonal phases of YMnO$_3$. The comparison of the experimental and first-principles calculated phonon spectra highlight unambiguously a spin-phonon coupling character in YMnO$_3$. This is further supported by the pronounced differences in the magnetic and non-magnetic phonon calculations. The calculated atomistic partial phonon contributions of the Y and Mn atoms are not affected by inclusion of magnetic interactions, whereas the dynamical contribution of the O atoms is found tochange. This highlights the role of the super-exchange interactions between the magnetic Mn cations, mediated by O bridges.  Phonon dispersion relations have also been calculated, in the entire Brillouin zone, for both the hexagonal phases. In the high-temperature phase, unstable phonon mode at the K point is highlighted. The displacement pattern at the K-point indicates that the freezing of this mode along with the stable mode at the Γ-point may lead to a stabilization of the low-temperature (P6$_3$cm) phase, and inducing ferroelectricity. Further, we have also estimated the mode Grüneisen parameter and volume thermal expansion behavior. The latter is found to agree with the available experimental data.

## I. INTRODUCTION

Manganites, $R$MnO$_3$ ($R$ = Dy–Lu, In, Y, and Sc), have been a subject of scientific interest for decades [1]. These materials belong to a distinguished class of multiferroics [2], since they exhibit ferroelectricity up to a very high temperature with the tripling of their unit cell [3]. The interplay between the related order parameters gives rise to remarkable features such as clamped ferroelectric-structural domain walls [4], with unusual transport properties [5]. Additionally, magnetoelectric properties [6-9] emerge due to the coupling of ferroelectricity with the low-temperature magnetic ordering [10]. Yttrium Manganese oxide (YMnO$_3$) keeps attracting a keen interest as it is known to exhibit ferroelectricity and antiferromagnetism simultaneously [10]. At ambient conditions the compound has a hexagonal P6$_3$cm structure. Above 1258 ±14 K, a ferroelectric to paraelectric phase transition occurs, and the system crystallizes [11] under the hexagonal P6$_3$/*mmc* space group (Fig. 1). Below 70 K, YMnO$_3$ has an A-type antiferromagnetic ordering. The ferroelectric phase consists of six formula units of YMnO$_3$. The structure is a framework network of MnO$_5$ bipyramids and YO$_7$ units. The MnO$_5$ units are tilted with respect to the c-axis and Y$^{3+}$ ions are shifted by ±δ from *a-b* plane along the c-axis. The high-temperature phase has two formula units of YMnO$_3$. The Mn$^{3+}$ ions are coordinated by five oxygens, whereas the Y$^{3+}$ ions are coordinated by six symmetrically equivalent oxygens, forming MnO$_5$ bipyramids and YO$_6$ units, respectively. All Y$^{3+}$ ions lie in the *a-b* plane. The ferroelectric P6$_3$cm phase is connected to the high-temperature *P6$_3$/mmc* structure by the tripling of the corresponding unit cell. Moreover there is a loss of the mirror symmetry perpendicular to the *c* axis due to the tilting and distortion of the MnO$_5$ bipyramids and the displacement of the Y atoms. This triggers the emergence of the spontaneous electric polarization of the system. The exact nature of this ferroelectric transition is still under debate [12–21]. Compounds [22–24] with reduced rare-earth ionic radius (Ho, Er, Tm, Yb, Lu and Y) crystallize in a hexagonal structure (space group P6$_3$cm), whereas an orthorhombic (space group *Pnma*) phase is reported for compounds with larger rare-earth ionic radius (La, Ce, Pr, Nd, Sm, Eu, Gd, Tb and Dy). A hexagonal-to-orthorhombic structural phase transition [25] can take place at elevated temperatures, under pressure.

Although ferroelectricity in YMnO$_3$ is believed to be due to the tripling of the unit cell, there is however an ambiguity about a second transition observed at ~ 920 K in this material. It was suggested [26] that this transition can be considered as a hidden order in which a residual symmetry, displayed by the trimerization order parameter, is spontaneously broken. In this context, a generic P6$_3$cm ↔ P3$c$1 ↔ *P*-3$c$1 phase diagram was proposed, corresponding to the observation in another isostructural compound (InMnO$_3$). High-resolution powder neutron diffraction investigation [11] of the structural behavior of the multiferroic hexagonal polymorph of YMnO$_3$ has been reported over the temperature range 300 - 1403 K. These measurements showed that on heating the ambient-temperature polar P6$_3$cm structure undergoes a centrosymmetric P6$_3$/mmc transition at 1258 ±14 K. This corroborated the absence of the previously suggested [17] intermediate phase with space group *P6$_3$/mcm*. Further, the measurements also provided evidence for an isosymmetric phase transition (i.e., *P6$_3$cm* to *P6$_3$cm*) at ≈ 920 K, which involves a sharp decrease in polarization.

Iliev et al [27] measured the Raman and infrared spectra of the hexagonal ferroelectric YMnO$_3$. The Raman and THz spectroscopic measurements pointed towards a strong spin-phonon coupling in YMnO$_3$ [28]. The inelastic neutron scattering (INS) measurements carried out on a single crystal indicated [29, 30] also a strong coupling between spins and phonons. Konstantin *et al* [31] performed first-principles calculations of phonon density of states and extracted the partial contributions of Y, Mn and O to the vibrational spectra. The authors have analyzed also the eigenvectors of the low-energy optic modes, and they calculated the sound velocity along various high symmetry directions from phonon dispersion relations. Further, *Alina et al* [32] performed zone centre phonon calculations of the ambient and high-temperature phases of YMnO$_3$. The structural phase transition of multiferroic YMnO$_3$ has been studied from first principles calculations [21] by performing detailed group theoretical analysis and density functional calculations of the total energy and phonons, highlighting unstable zone boundary mode which freeze with temperature, and a coupling with a zone centre mode. The latter contributing to emergence of the polarization in the compound. This coupling is the mechanism that allows the emergence of multiferroicity in this class of materials. It was concluded that YMnO$_3$ is an improper ferroelectric. Various theoretical and experimental attempts have been made to understand the anomalous properties of YMnO3 as well as to understand the mechanism of phase transitions, multiferroicity, structure and dynamics of the compounds [33-44].

Most of the dynamical probes of manganites are based on Raman and infrared techniques [27, 28]. These are limited to the zone centre modes. Available INS measurements on single crystals are limited to low energies, below 20 meV [29, 30]. However a better understanding of thermodynamical behavior of a material requires a complete description of phonon spectra in the entire Brillouin zone (zone-centre and zone-boundary). Therefore INS can be used to probe phonons in the entire Brillouin zone. In this context, we present temperature-dependent measurements of phonon spectra in YMnO$_3$, over the temperature range 50 - 1303 K .i.e. in the low-temperature (*P6$_3$cm*), as well as in the high-temperature (P6$_3$/*mmc*) phase of YMnO$_3$. Our measurements are accompanied by first principles density functional-based magnetic lattice dynamics calculations for the sake of analysis and interpretation of the neutron data. Phonon dispersion relations in the entire Brillouin zone have been calculated in both the high- and low-temperature hexagonal phases of YMnO$_3$. Unstable phonon modes in the high-temperature phase are related to the stabilization of the low-temperature phase. This paper is organized as follows: the experimental and computational techniques are summarized in section II and section III, respectively. Section IV is dedicated to the presentation and discussion of the results. Conclusions are drawn in section V.

## II. EXPERIMENTAL DETAILS

The inelastic neutron scattering measurements were carried out, at several temperatures in the range 70 - 1303 K, using the direct-geometry thermal neutron time-of-flight IN4C spectrometer at the Institut Laue-Langevin (ILL), France. The spectrometer is equipped with a large detector bank covering a wide scattering angle range of about 10$^o$ to 110$^o$. About 10 grams of polycrystalline sample of YMnO$_3$, prepared by solid state reaction method [45], has been used for the measurements. The low-temperature measurements were performed using a standard orange cryostat. For the high-temperature range, the sample was put into a quartz tube, and mounted in a furnace. The other end of the quartz tube was kept open in the air. An incident neutron wavelength of 2.4 Å (14.2 meV) was used, performing in the up-scattering mode (neutron energy gain). The momentum transfer, Q, extends up to 7 Å$^{-1}$. In the incoherent one-phonon approximation, the measured scattering function $S(Q,E)$, as observed in the neutron experiments, is related [46] to the phonon density of states $g^{(n)}(E)$ as follows:

$$g^{(n)}(E) = A < \frac{e^{2W_k(Q)}}{Q^2} \frac{E}{n(E,T) + \frac{1}{2} \pm \frac{1}{2}} S(Q,E) > \qquad (1)$$

$$g^n(E) = B \sum_k \{ \frac{4\pi b_k^2}{m_k} \} g_k(E) \qquad (2)$$

where the + or − signs correspond to energy loss or gain of the neutrons respectively and where $n(E,T) = \left[\exp(E/k_BT) - 1\right]^{-1}$. $A$ and $B$ are normalization constants and $b_k$, $m_k$, and $g_k(E)$ are, respectively, the neutron scattering length, mass, and partial density of states of the $k^{th}$ atom in the unit cell. The quantity between < > represents suitable average over all $Q$ values at a given energy. $2W(Q)$ is the Debye-Waller factor. The weighting factors $\frac{4\pi b_k^2}{m_k}$ for various atoms in the units of barns/amu are:

Y: 0.0866; Mn: 0.0385 and O: 0.2645. The values of the neutron scattering lengths are taken from Ref. [47]. The multiphonon contribution has been calculated using the Sjølander formalism [46(b),48] and subtracted from the experimental data.

## III. COMPUTATIONAL DETAILS

Relaxed geometries and total energies were obtained using the projector-augmented wave (PAW) formalism [49, 50] of the Kohn-Sham density functional theory [51, 52], within both the local density approximation (LDA) and the generalized gradient approximation (GGA), implemented in the Vienna ab-initio simulation package (VASP) [53]. The GGA was formulated by the Perdew–Burke–Ernzerhof (PBE) density functional [54]. The LDA was based on the Ceperly–Alder parametrization by Perdew and Zunger [55]. The valence electronic configurations of Y, Mn and O as used in calculations for pseudo potential generation are $4s^24p^65s^24d^1$, $d^6s^1$ and $s^2p^4$, respectively. The calculations were carried out with and without considering the magnetic interactions. This is for the sake of understanding the effect of the spin degrees of freedom on the lattice dynamics [56-58]. The A-type antiferromagnetic ordering, in both the hexagonal phases, was adopted in the magnetic calculations. The on-site Coulomb interaction was accounted for within the Dudarev approach [59] using $U_{\text{eff}} = U - J = 7.12$ eV [60, 61].

Both full (lattice parameters and atomic positions) and partial (only atomic positions) geometry relaxations were carried out. Hereafter, the labeling "FM" and "FNM" refers to fully relaxed magnetic and fully relaxed non-magnetic calculations, respectively. Further, "PNM" refers to the partially relaxed non-magnetic calculation, where we used the structure obtained from "FM" and relaxed only the atomic positions without magnetic ordering. An energy cutoff of 800 eV and a 6×6×2 k-points mesh for the Brillouin zone integration are used and found to satisfy the required numerical convergence. All results are well converged with respect to k-mesh and energy cutoff for the plane wave expansion. The break conditions for the self-consistent field and for the ionic relaxation loops were set to $10^{-8}$ eV and $10^{-5}$ eV Å$^{-1}$, respectively. The later criteria means that the obtained Hellmann–Feynman forces are less than $10^{-5}$ eV.Å$^{-1}$. Total energies were calculated for 24 generated structures resulting from individual displacements of the symmetry inequivalent atoms in the room temperature hexagonal (P6$_3$cm), and high temperature (P6$_3$/mmc) phases, along the six inequivalent Cartesian directions (±x, ±y and ±z). Phonons are extracted from subsequent calculations using the direct method as implemented in the Phonon software [62].

## IV. RESULTS AND DISCUSSION

### A. Temperature dependence of phonon spectra

The INS measurements are performed across the magnetic and structural transition temperatures. In addition to the temperature dependence of the INS spectra, we present the data in terms of the Q-dependence as well, in order to map out both magnetic and lattice (phonon) excitations. The magnetic scattering intensity increases as the temperature and Q decrease, whereas a phonon signal exhibits the reverse trend. Hereafter the Q-dependence is limited to compare INS spectra in the low-Q (1 - 4 Å$^{-1}$) and high-Q (4 - 7 Å$^{-1}$) regions.

The measured temperature dependence of the dynamical scattering function $S(Q,\omega)$ within the low-Q and high-Q regions is shown in Fig. 2(a) and Fig 2(b), respectively. At 50 K the observed peak around 10 meV is of magnetic nature, given its Q-dependence (stronger at low-Q and absent at high-Q). Above 70 K the compound undergoes [10, 63] a paramagnetic transition. Our data at 150 and 315 K show an appreciable broadening of the elastic peak which may results from the paramagnetic scattering. The increase in the temperature would lead to a decay of the paramagnetic fluctuations, which reduces the width of the elastic peak. Above 848 K the width of the elastic line stays close to the instrumental resolution, indicating that the paramagnetic scattering vanishes at high temperatures.

The neutron inelastic spectra are depicted in Fig. 2(c) and Fig. 2(d), for the low-Q and high-Q regions, respectively. For the low-Q data, the low-energy excitations below 20 meV show a pronounced change as a function of the temperature. However such a trend is not observed in the high-Q data, confirming a more phonon-like character in this spectral range. Both the magnetic and phonon excitations are present in the low-Q part, whereas the high-Q data contains mainly phonon contributions. A prominent peak is observed in the low-Q data of the neutron inelastic spectra at 50 K and decaying strongly with temperature. The INS measurements [29, 30] carried out on a single crystal of YMnO$_3$ indicate that the dispersion of spin waves lies within the energy range 5 - 15 meV. This is in agreement with our measurements where we also find a large change in the intensity in the low-Q data, across the Neel temperature (T$_N$~70 K). The stretching modes (Fig. 2) around 80 meV, at 315 K, are found to soften as the temperature increases to 848 K. This might primarily be due to the increase of the Mn-O bond lengths leading to such a softening.

The structural phase transition (P6$_3$cm to P6$_3$/mmc) in YMnO$_3$, occurring at 1258 ±14 K, involves a structural distortion and a ferroelectric to paraelectric transition. It is well know that the phase transition from ferroelectric to paraelectric is driven by a softening of a zone boundary phonon mode at the K-point. We did not observe any significant change across the transition (up to 1303 K). It should be noted that the neutron inelastic spectra measurements were performed on a polycrystalline sample, and are averaged over the whole Brillouin zone. Therefore it would be crucial to detect small changes in specific dispersion branches, which may only be measured efficiently using a single crystal.

**B. Effect of the magnetic ordering on the calculated phonon spectra**

The purpose of the three model calculations ("FM", "FNM", and "PNM") described in Section III is to study the effect of the cell volume and magnetic interactions on the phonon spectra. The comparison between the relaxed and the experimentally refined structural parameters are gathered in TABLE I. Further, a comparison between calculated and refined [11] bond lengths is provided in TABLE II. The calculated magnetic moment per Mn atom is 4.05 $\mu_B$ and 4.2 $\mu_B$, from LDA and GGA, respectively. However neutron diffraction measurements [64, 65] reported that, below T$_N$, the Mn moment has a value of ~ 3 $\mu_B$. YMnO$_3$ is a non collinear magnetically frustrated two dimensional system [3]. The deviation from the calculated value might be due to magnetic fluctuations associated with frustration and/or low dimensionality.

The calculated phonon spectra are compared with the measured ones (at 315 K) in Fig. 3. The "FM" type calculation using GGA reproduces the low energy features of the measured phonon spectra. The peak in the experimental spectra at about 52 meV is estimated around 45 meV. Further the Mn-O stretching modes are underestimated around 67 meV, while experimentally these are observed around 80 meV. The GGA determined structural parameters clearly show that both the 'a' and 'c' lattice parameters are overestimated by ~ 2 %. This results in the overestimation of the various Mn-O bond lengths, which leads to the underestimation of the energies of the Mn-O stretching modes. Results of the FM-LDA calculations are found to be close to the experimental data. This is in agreement with the correct estimation of the structural parameters as well as bond lengths using this model calculation (TABLE I and II). Consequently in the following we adopt the LDA density functional.

The calculated phonon spectra (Fig. 3), using the three model calculations PNM, FM, FNM, are closely similar up to ~ 25 meV. A significant change is observed at higher energies. To gain deeper insights into the phonon spectra, we extracted the atomistic partial contributions to the calculated phonon density of states (Fig. 4). The Y and Mn atoms are found to contribute mainly below 50 meV, while the O atoms are dynamically active within the range 40 - 90 meV. In all the three model calculations the contributions from Y and Mn atoms remain nearly unperturbed in the entire energy range. On the other hand, the differences in the phonon spectra are found to originate from the dynamics of the oxygen atoms. Such differences from the various calculations are primarily due to the nature of the chemical bonding, in the magnetic and nonmagnetic configurations, as well as a volume effect in this case as well. By comparing the FNM and FM phonon calculations we highlighted an effect of magnetism and cell volume on the phonon spectra.

By comparing the FM and PNM calculated phonon spectra we can identify the specific modes sensitive to magnetic interactions. In both FM and PNM, the calculated lattice parameters (TABLE I) are similar. However there are slight changes in the related atomic positions. We find that the calculated Mn-O1 and Mn-O2 bond lengths show a large difference in both the calculations (TABLE II). The FM calculated bond lengths are closer to the experimentally refined ones. The calculated values of Mn-O2 are 1.879 Å and 1.806 Å in "FM" and "PNM", respectively. The significant shortening of the Mn-O2 bond length as compared to the refined value of 1.891 Å [11], results in a hardening of the modes (Fig. 4) around 80 meV in the "PNM" calculations. The overestimation (TABLE II) of Mn-O1 bond lengths in the "PNM" calculations leads to a softening of the modes involving O atoms, in comparison with "FM" calculations. Further, we find that the PNM-calculated low-energy modes around 40 meV for the planar O3 and O4 atoms soften significantly in comparison (Fig. 4) to the "FM" calculations. The difference in the calculated phonon spectra in both FM and PNM is due to the fact that in the PNM calculations the $Mn^{3+}$ magnetic moment is zero, while the FM calculated magnetic moment of $Mn^{3+}$ is 4.05 $\mu_B$. Therefore, the super exchange interactions between Mn cations, mediated by O atoms, influence the partial phonon spectra involving oxygens. It it worth to notice that the Mn phonon modes seems to be insensitive to the magnetic moment exhibited on the Mn site.

The FNM calculated structure (TABLE II) shows that the "c" lattice parameter has a value of 12.01 Å, while the experimentally refined value is 11.40 Å [11]. The "a" lattice parameter is found to be underestimated (5.838 Å) in comparison to the experimental value of 6.1415 Å. All the four FNM calculated Mn-O bond lengths (TABLE II) are more isotropic. This leads to an underestimation of the energies of the Mn-O stretching modes (Fig. 4) around 80 meV in the "FNM" calculations in comparison to the "FM" calculations.

Fig. 5 shows the change in the energy of the estimated zone centre modes in the PNM and FNM configuration, with respect to FM calculations. There is a noticeable deviation when comparing energies from magnetic and non-magnetic calculations. The maximum shift in phonon energies is observed for high energy Mn-O stretching modes. This supports a spin-phonon coupling behavior. The change in the energies of the modes around 30 meV is mainly due to magnetic interactions, while the high energy stretching modes are most influenced by the estimated Mn-O bond lengths.

**C. Phonon spectra in the room temperature (P6$_3$cm) and high-temperature (P6$_3$/mmc) hexagonal phases**

The calculated phonon spectra of YMnO$_3$ were subject of previous works [21, 31, 66]. The estimated phonon dispersions have been reported [31] in the entire Brillouin zone, at room temperature in the P6$_3$cm hexagonal phase. However, only calculations of phonon modes at the zone centre and K-point were reported [21] in the high-temperature hexagonal phase (P6$_3$/mmc). The group theoretical analysis along with the first-principles phonons calculations at zone centre is used to understand [21] the mechanism of multiferroicity in YMnO$_3$. Alina et al. [32] calculated the zone-centre phonon modes using LDA, and they reassigned the Raman modes. We went beyond obtaining only the zone-centre modes by extracting also phonon dispersion relations (Fig. 6), in the entire Brillouin zone and along various high-symmetry directions in both the hexagonal phases (P6$_3$cm and P6$_3$/mmc). We discuss also the relationship between the phonon modes at the ferroelectric (*P6$_3$cm*) to the paraelectric (*P6$_3$/mmc*) phase transition.

The calculated phonon dispersion relations in both the room-temperature (space group P6$_3$cm) and the high-temperature (space group P6$_3$/mmc) phases are shown in Fig. 6. The phonon modes are found to be stable in the entire Brillouin zone, in the low-temperature phase. However, phonon instability is clearly noticed in the high-temperature phase, at the symmetry point K (1/3 1/3 0). The unstable mode is highly anharmonic in nature, and it become stable at higher temperatures due to anharmonicity. It has been proposed [8, 12] that the condensation of the unstable phonon mode at K point drives the transition to the low-temperature structure of YMnO$_3$. This mode is not polar in nature. However, ferroelectricity in the improper ferroelectric YMnO$_3$ arises from the coupling of the unstable K-point mode with a stable mode at the Γ-point. The latter is polar in nature and, therefore, contributing to the ferroelectricity in the room-temperature phase. The eigenvectors of these modes have been extracted from our ab-initio calculations. The displacement pattern of unstable mode at the K point is shown in Fig 7. At the K-point, the mode consists of an unequal displacement of two Y atoms in opposite direction, along with an out-of-phase rotation of MnO$_5$ bipyramid units, around the c-axis. The unequal amplitude of motions of the O atoms induces a distortion of the MnO$_5$ units. The displacement pattern of the stable mode at the Γ-point consists of a motion of O atoms belonging to the plane formed by the Mn atoms of the MnO$_5$ units.

We have also calculated the phonon dispersion relation of high temperature phase (P6$_3$/*mmc*) of YMnO$_3$ (Fig. 6) with a super cell of √3×√3×1 which is equivalent to room temperature hexagonal phase (P6$_3$*cm*). The structure as used in the phonon calculations is given in TABLE III. The comparison of the two structures show that in the room temperature phase (P6$_3$*cm*) atomic positions are slightly distorted in comparison to the structure obtained from the √3×√3×1 super cell of the high temperature phase (P6$_3$/*mmc*). The group theoretical classification at zone centre of low temperature phase (P6$_3$*cm*) is 10A$_1$+5A$_2$+5B$_1$+10B$_2$+30E$_1$+30E$_2$ while the classification at K point in the high temperature phase (P6$_3$/*mmc*) is given by 2K$_1$+2K$_2$+3K$_3$+3K$_4$+12K$_5$+8K$_6$. The freezing of unstable K point mode in the high temperature phase will lead to transition to the room temperature phase. Group theoretical analysis shows that the unstable mode at K point belongs to K$_3$ representation and condenses to stable modes of A1 and B1 representations in room temperature phase (P6$_3$*cm*). The difference in the atomic co-ordinates of room temperature phase (P6$_3$*cm*) and the √3×√3×1 super cell of the high temperature phase (P6$_3$/*mmc*) is a measure of the distortion required to stabilize the ambient temperature phase. The eigen vectors of the unstable K3 mode of high temperature phase (P6$_3$/mmc) for the super cell is given in Table III. The eigen vector of the K3 mode is in good agreement with the distortion vector, which corroborates the previous results [11, 21] that freezing of the unstable K3 mode is responsible for phase transition to the ambient temperature phase.

**D. Thermal expansion Behavior**

The calculation of the thermal expansion of YMnO$_3$ is carried out within the quasi-harmonic approximation (QHA). In QHA, each phonon mode contributes to the volume thermal expansion coefficient [67, 68], given by: $\alpha_V = \frac{1}{BV}\sum_i \Gamma_i C_{Vi}(T)$, with $\Gamma_i(=-\partial lnE_i/\partial lnV)$ and C$_{vi}$ are the mode Grüneisen parameter and the specific heat of the i$^{th}$ vibrational state of the crystal, respectively. The volume dependence of phonon modes is calculated in the entire Brillouin zone. The pressure dependence of the phonon spectra in the entire Brillouin zone was extracted from LDA-based "FM" calculations, at two pressure points: ambient and 0.5 GPa. The calculated mode Grüneisen parameter values Γ(E) (Fig. 8(a)) depend strongly on the mode energy, and are found to lie within 0.2 - 4.0. The thermal expansion behaviour (Fig. 8(b)) has been calculated up to 1250 K, and compared to available experimental data [69,11]. Our simulated volume thermal expansion agrees well with the observed behavior up to 700 K. At higher temperatures the estimated thermal expansion evolution is underestimated with respect to experimental data. Such an underestimation could be due to the fact that we have considered only the implicit part of the anharmonicity in our calculations. This would points towards a larger effect of the explicit component of the anharmonicity in describing the total thermal expansion behavior, due to the large thermal amplitude.

## V. CONCLUSIONS

We have reported measurements of neutron inelastic scattering spectra of the multiferroic material YMnO$_3$ over a wide temperature range (50-1303 K) covering all the relevant characteristic transition temperatures. The room temperature phase is found to be subject to a strong spin-phonon coupling. The calculated phonon dispersion relations in the entire Brillouin zone indicate phonon instability in the high-temperature (P6$_3$/*mmc*) hexagonal phases of YMnO$_3$, at the symmetry point K (1/3 1/3 0). Unstable phonon modes may lead to a stabilization of the low-temperature (P6$_3$*cm*) phase. Further, the pressure dependence of phonon energies in the entire Brillouin zone is used to determine the thermal expansion behavior of the room temperature phase of YMnO$_3$ (P6$_3$*cm*), and which is found to be in a good agreement with the available experimental data.

TABLE I. Comparison between the calculated and room temperature experimental [11] structural parameters of YMnO$_3$ (Hexagonal phase, space group P6$_3$cm). The experimental structure (space group P6$_3$cm) consists of Y1 and O3 atoms at *2a(x,y,z)*, Y2 and O4 at *4b(x,y,z)*, and Mn, O1, and O2 at *6c(x,y,z)* Wyckoff site. "FM", "FNM" and "PNM" refer to fully relaxed magnetic, fully relaxed non-magnetic and partially relaxed non magnetic calculations, respectively.

|     |     | **Expt.** | **FM(GGA)** | **FM(LDA)** | **PNM(LDA)** | **FNM(LDA)** |
|-----|-----|-----------|-------------|-------------|--------------|--------------|
|     | *a (Å)* | 6.14151 | 6.253 | 6.095 | 6.095 | 5.838 |
|     | *b (Å)* | 6.14151 | 6.253 | 6.095 | 6.095 | 5.838 |
|     | *c (Å)* | 11.4013 | 11.644 | 11.416 | 11.416 | 12.013 |
| O1  | *x* | 0.3074 | 0.308 | 0.306 | 0.307 | 0.302 |
|     | *y* | 0.0000 | 0.000 | 0.000 | 0.000 | 0.000 |
|     | *z* | 0.1626 | 0.165 | 0.164 | 0.169 | 0.161 |
| O2  | *x* | 0.6427 | 0.645 | 0.640 | 0.640 | 0.636 |
|     | *y* | 0.0000 | 0.000 | 0.000 | 0.000 | 0.000 |
|     | *z* | 0.3355 | 0.335 | 0.336 | 0.342 | 0.339 |
| O3  | *x* | 0.000 | 0.000 | 0.000 | 0.0000 | 0.0000 |
|     | *y* | 0.000 | 0.000 | 0.000 | 0.0000 | 0.0000 |
|     | *z* | 0.4744 | 0.475 | 0.476 | 0.4876 | 0.475 |
| O4  | *x* | 0.3333 | 0.333 | 0.333 | 0.333 | 0.333 |
|     | *y* | 0.6667 | 0.667 | 0.667 | 0.667 | 0.667 |
|     | *z* | 0.0169 | 0.021 | 0.021 | 0.028 | 0.020 |
| Mn  | *x* | 0.3177 | 0.333 | 0.333 | 0.336 | 0.333 |
|     | *y* | 0.0000 | 0.000 | 0.000 | 0.000 | 0.000 |
|     | *z* | 0.0000 | 0.000 | 0.000 | 0.000 | 0.000 |
| Y1  | *x* | 0.0000 | 0.000 | 0.000 | 0.000 | 0.000 |
|     | *y* | 0.0000 | 0.000 | 0.000 | 0.000 | 0.000 |
|     | *z* | 0.2728 | 0.274 | 0.275 | 0.276 | 0.274 |
| Y2  | *x* | 0.6667 | 0.667 | 0.667 | 0.667 | 0.667 |
|     | *y* | 0.3333 | 0.333 | 0.333 | 0.333 | 0.333 |
|     | *z* | 0.7325 | 0.732 | 0.731 | 0.731 | 0.733 |

TABLE II. Comparison between the calculated and room temperature experimental [11] (293 K) bond lengths (in Å units) in YMnO$_3$ (Hexagonal phase, space group P6$_3$cm). "FM", "FNM" and "PNM" refer to fully relaxed magnetic, fully relaxed non-magnetic and partially relaxed non magnetic calculations, respectively.

| Bond | Expt. | GGA (FM) | LDA(FM) | LDA(PNM) | LDA(FNM) |
|---|---|---|---|---|---|
| Mn-O1 | 1.855 | 1.931 | 1.881 | 1.940 | 1.934 |
| Mn-O2 | 1.891 | 1.929 | 1.880 | 1.806 | 1.939 |
| Mn-O3 | 1.973 | 2.105 | 2.051 | 2.051 | 1.968 |
| Mn-O4 | 2.106 | 2.098 | 2.045 | 2.049 | 1.961 |
| Y1-O1 | 2.268 | 2.305 | 2.255 | 2.232 | 2.244 |
| Y1-O2 | 2.308 | 2.350 | 2.302 | 2.322 | 2.256 |
| Y1-O3 | 2.299 | 2.341 | 2.290 | 2.418 | 2.389 |
| Y2-O1 | 2.275 | 2.303 | 2.255 | 2.231 | 2.220 |
| Y2-O2 | 2.300 | 2.341 | 2.292 | 2.331 | 2.261 |
| Y2-O4 | 2.458 | 2.458 | 2.403 | 2.323 | 2.556 |
| Mn-Mn | 3.632 | 3.609 | 3.519 | 3.507 | 3.372 |
| Mn-Y1 | 3.243 | 3.357 | 3.272 | 3.277 | 3.320 |
| Mn-Y2 | 3.701 | 3.411 | 3.681 | 3.677 | 3.755 |
| Y1-Y1 | 5.701 | 5.822 | 5.708 | 5.708 | 6.006 |
| Y1-Y2 | 3.575 | 3.643 | 3.555 | 3.556 | 3.410 |
| Y2-Y2 | 3.546 | 3.610 | 3.519 | 3.519 | 3.371 |

TABLE III. The calculated structures in the ambient temperature and super cell ($\sqrt{3}\times\sqrt{3}\times1$) of high temperature phase. The super cell ($\sqrt{3}\times\sqrt{3}\times1$) of high temperature phase is equivalent to the room temperature hexagonal phase (P6$_3$cm). The unit cell in the space group P6$_3$cm have Y1 and O3 atoms at the *2a(x,y,z)*, Y2 and O4 at *4b(x,y,z),* and Mn, O1, and O2 at *6c(x,y,z)* Wyckoff site. The distortion vector is obtained from the difference in atomic co-ordinates of the ambient temperature (P6$_3$cm) and super cell of high temperature phases. The eigen vector of the unstable K3 mode in the high temperature phase (P6$_3$/*mmc*) for the super cell is also given. The amplitude of the eigen vector of O1 is scaled to match with the distortion vector.

|  |  | Ambient temperature phase (*P6$_3$cm*) | Supercell of high temperature phase (*P6$_3$/mmc*) | Distortion vector in fractional coordinates | Eigen vector of the unstable K3 mode in the high temperature phase (*P6$_3$/mmc*) for the super cell in fractional coordinates |
|---|---|---|---|---|---|
|  | *a (Å)* | 6.095 | 6.165 |  |  |
|  | *b (Å)* | 6.095 | 6.165 |  |  |
|  | *c (Å)* | 11.416 | 11.223 |  |  |
|  |  |  |  |  |  |
| O1 | *x* | 0.306 | 0.333 | -0.027 | -0.027 |
|  | *y* | 0.000 | 0.000 | 0.000 | 0.000 |
|  | *z* | 0.164 | 0.1665 | -0.003 | 0.000 |
|  |  |  |  |  |  |
| O2 | *x* | 0.640 | 0.667 | -0.027 | -0.027 |
|  | *y* | 0.000 | 0.000 | 0.000 | 0.000 |
|  | *z* | 0.336 | 0.334 | 0.002 | 0.000 |
|  |  |  |  |  |  |
| O3 | *x* | 0.000 | 0.000 | 0.000 | 0.000 |
|  | *y* | 0.000 | 0.000 | 0.000 | 0.000 |
|  | *z* | 0.476 | 0.500 | -0.024 | -0.039 |
|  |  |  |  |  |  |
| O4 | *x* | 0.333 | 0.333 | 0.000 | 0.000 |
|  | *y* | 0.667 | 0.667 | 0.000 | 0.000 |
|  | *z* | 0.021 | 0.000 | 0.021 | 0.020 |
|  |  |  |  |  |  |
| Mn | *x* | 0.333 | 0.333 | 0.000 | 0.000 |
|  | *y* | 0.000 | 0.000 | 0.000 | 0.000 |
|  | *z* | 0.000 | 0.000 | 0.000 | 0.000 |
|  |  |  |  |  |  |
| Y1 | *x* | 0.000 | 0.000 | 0.000 | 0.000 |
|  | *y* | 0.000 | 0.000 | 0.000 | 0.000 |
|  | *z* | 0.275 | 0.250 | 0.025 | 0.019 |
|  |  |  |  |  |  |
| Y2 | *x* | 0.667 | 0.667 | 0.000 | 0.000 |
|  | *y* | 0.333 | 0.333 | 0.000 | 0.000 |
|  | *z* | 0.731 | 0.750 | -0.019 | -0.018 |

FIG. 1 (Color online) Schematic representation of the crystal structure of the room-temperature (space group P6$_3$cm) and the high-temperature (space group P6$_3$/*mmc*) phases of YMnO$_3$. The atoms are labeled following Table I. Key: Y, blue spheres; Mn, green spheres; and O, red spheres.

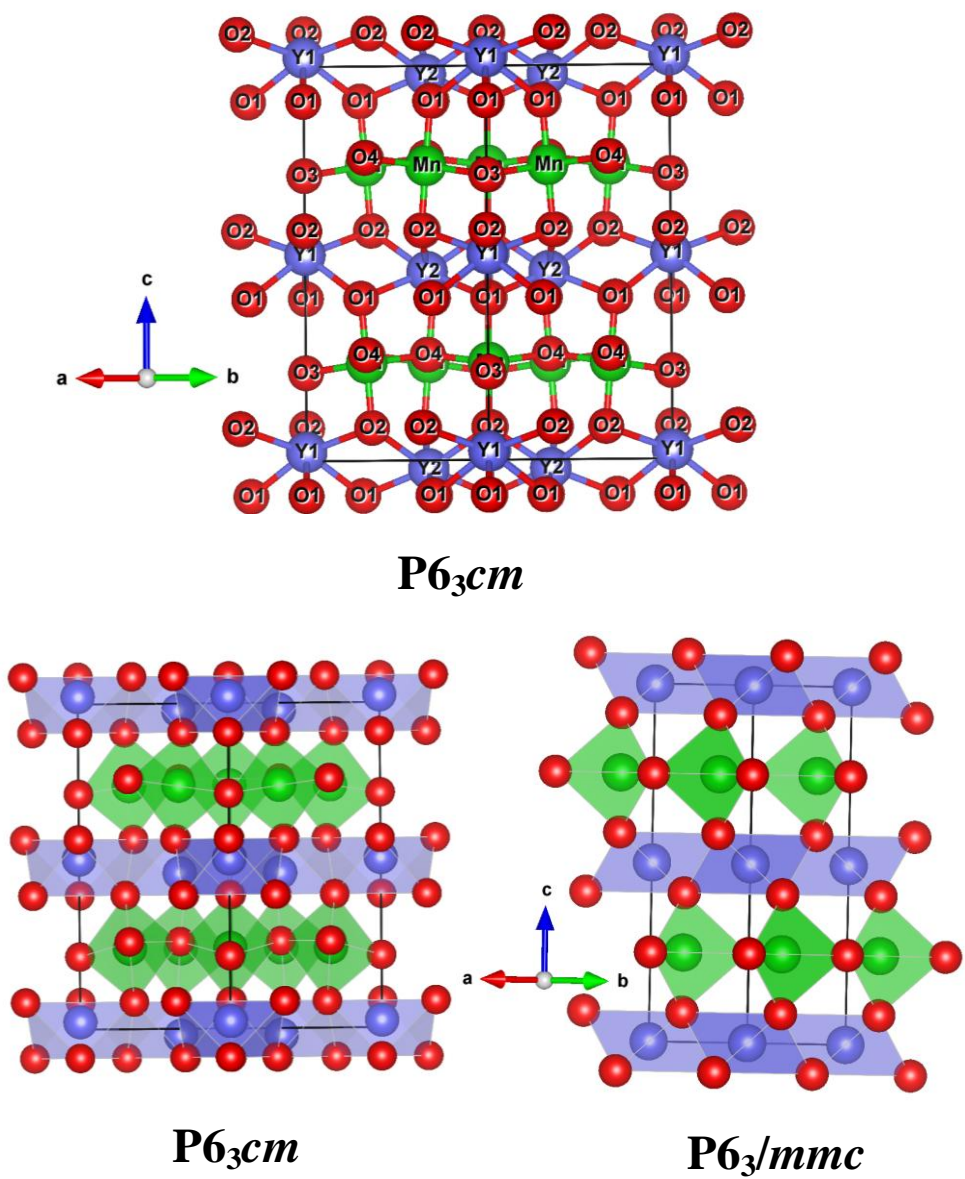

**P6$_3$cm**

**P6$_3$cm**              **P6$_3$/*mmc***

FIG. 2 (Color online) Temperature dependent neutron inelastic spectra of YMnO$_3$. Top panel: the low-Q and high-Q Bose factor corrected S(Q,E), where both the energy loss (0 - 10 meV) and the energy gain (-20 - 0 meV) sides are shown. Bottom panel: the low-Q and high-Q, unity-normalized, neutron inelastic spectra, g$^{(n)}$(E), inferred from the neutron energy gain mode S(Q,E) data, within the incoherent approximation.

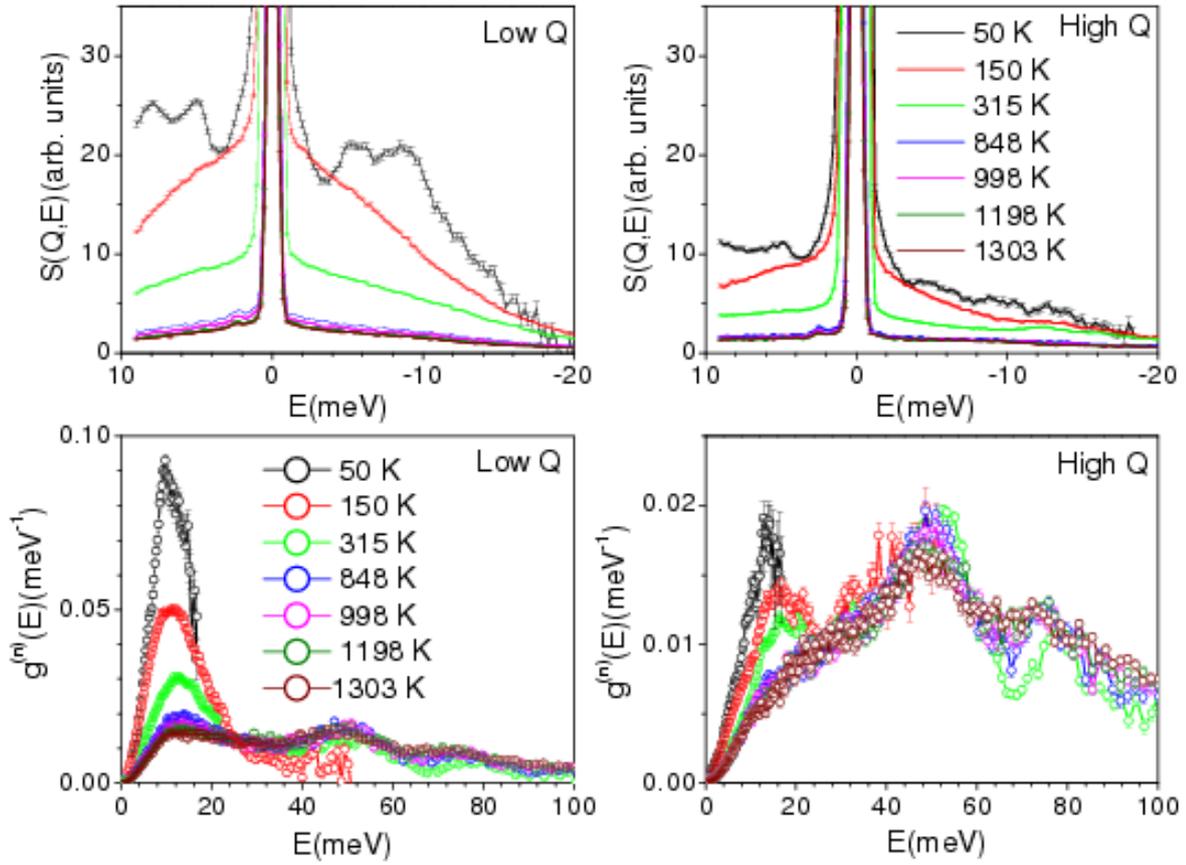

FIG. 3 (Color online) The calculated and experimental neutron inelastic spectra of YMnO$_3$. The experimental data were collected at 315 K, and averaged over the high-Q region. The calculated phonon spectra have been convoluted with a Gaussian of FWHM of 15% of the energy transfer in order to describe the effect of energy resolution in the experiment. For better visibility, the experimental and calculated phonon spectra are shifted vertically with respect to each other. Multiphonon as calculated using the Sjølander formalism has been subtracted for comparison with the calculations

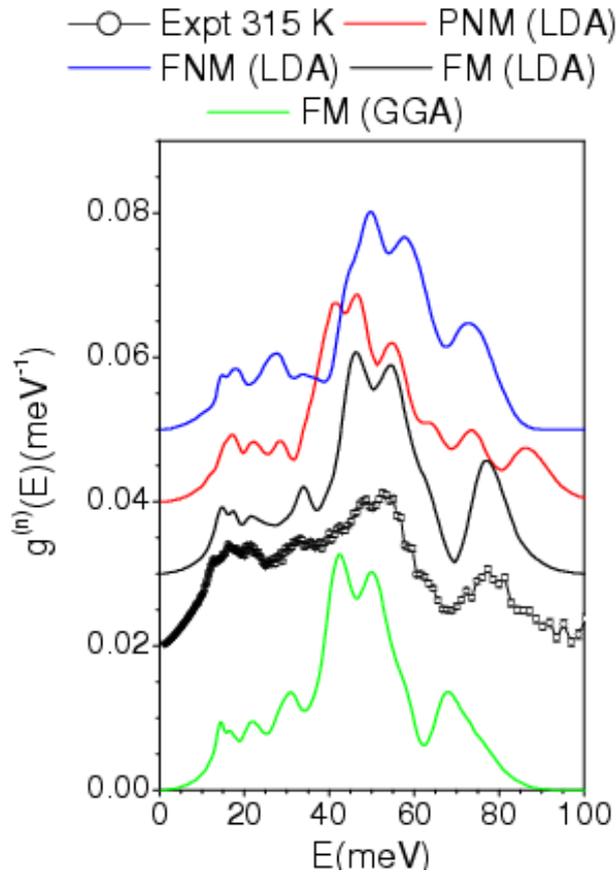

FIG. 4 (Color online) The calculated atomistic partial phonon density of states (Y, Mn and O) in the low temperature phase (space group P6$_3$cm) of YMnO$_3$, within the local density approximation (LDA). The atoms are labeled following Table I. "FM", "FNM" and "PNM" refer to fully relaxed magnetic, fully relaxed non-magnetic and partially relaxed non magnetic calculations, respectively.

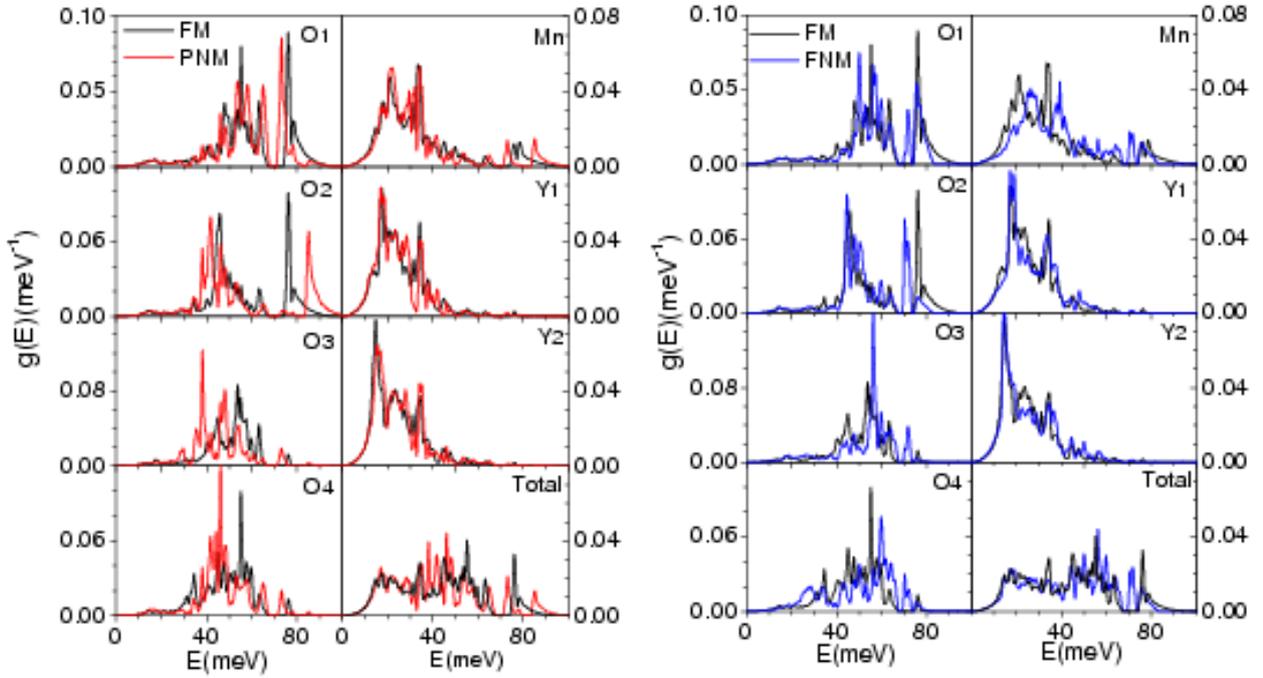

FIG. 5 (Color online) The calculated shift of the zone centre optic phonon modes in "PNM" and "FNM" configurations with respect to the "FM" model calculation. The calculations are performed in the low temperature phase (space group P6$_3$cm) of YMnO$_3$, within the local density approximation (LDA).

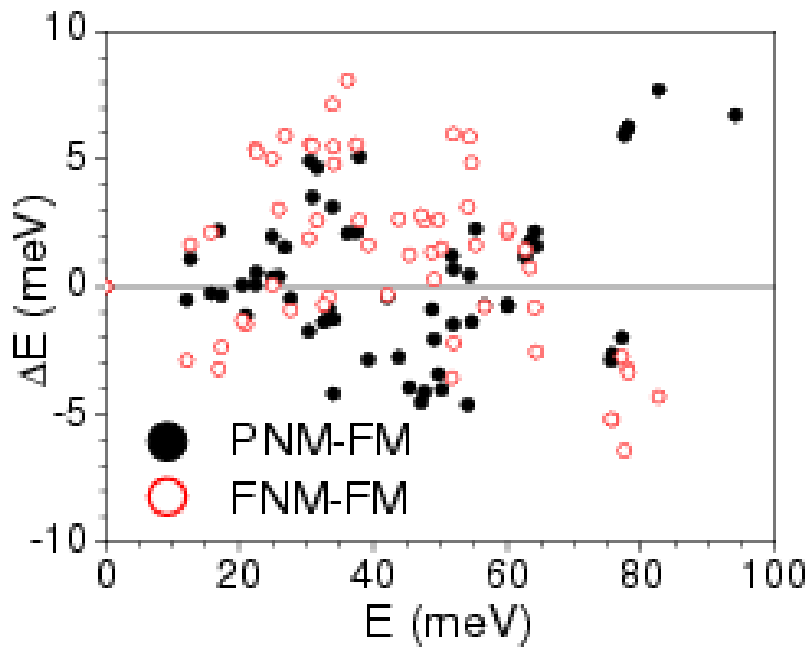

FIG. 6 (Color online) The calculated phonon dispersion relations along the high-symmetry directions of the ambient-temperature (space group P6$_3$cm) and the high-temperature (space group P6$_3$/*mmc*) hexagonal phases of YMnO$_3$, within the local density approximation (LDA). The zoom of the calculated phonon dispersion relations in the ambient temperature (P6$_3$cm) and high temperature phase (P6$_3$/*mmc*) with a super cell of √3×√3×1 are also shown. The high-symmetry points are: K (1/3 1/3 0), A (0 0 1/2) and Γ (0 0 0). The (1/2/1/2 0) and (1/2 0 0) are equivalent M point in the hexagonal symmetry. The size of the hexagonal unit cell is nearly same in the P6$_3$cm and super cell of P6$_3$/*mmc*. The special points (A, K, M, Γ) shown for the super cell of P6$_3$/*mmc* are those of the space group P6$_3$cm.

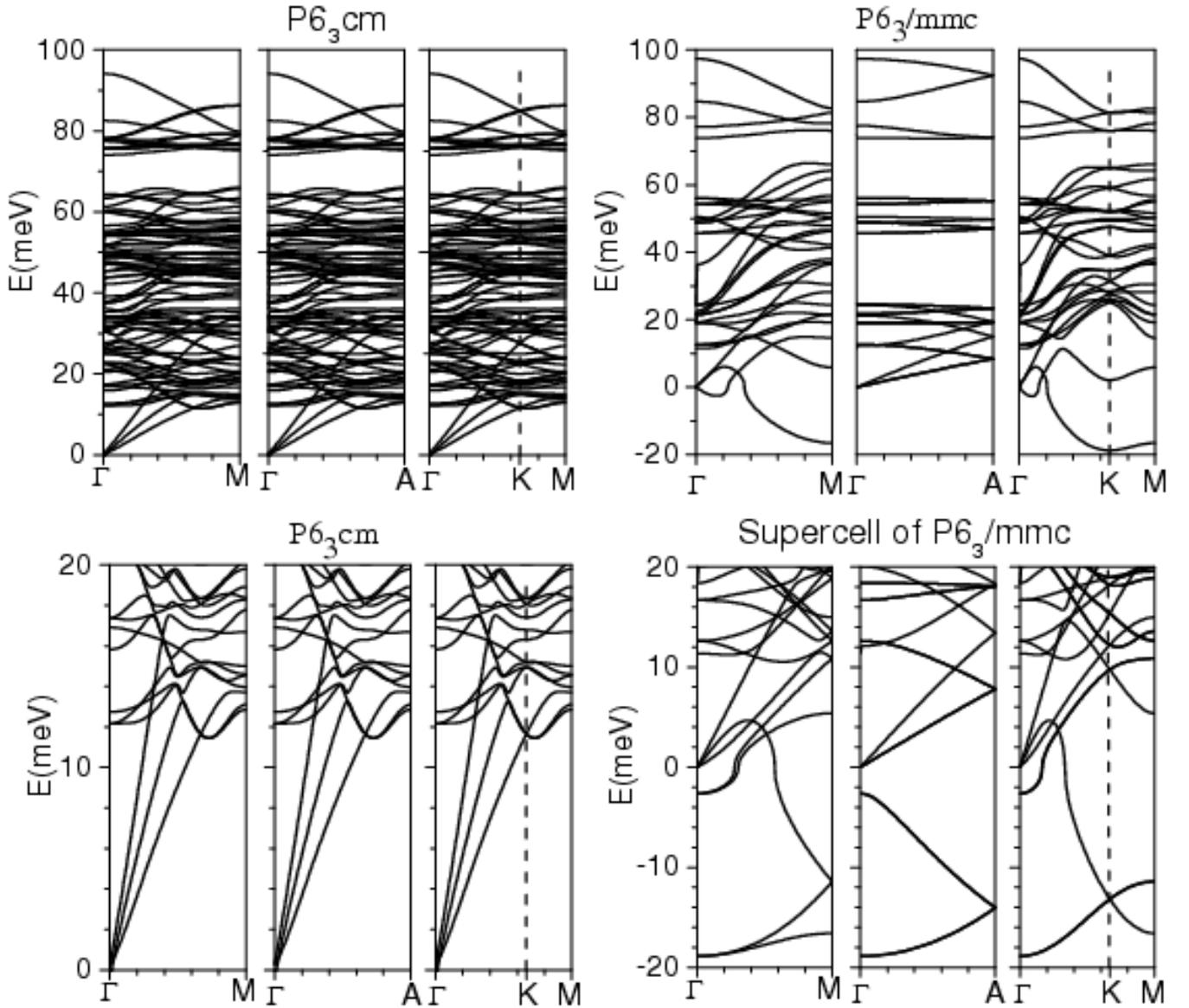

FIG. 7 (Color online) The displacement patterns of the lowest phone mode at K and Γ points in the high-temperature phase (space group P6$_3$/*mmc*) of YMnO$_3$. The lengths of arrows are related to the displacements of the atoms. The absence of an arrow on an atom indicates that the atom is at rest. Key: Y, blue spheres; Mn, green spheres; and O, red spheres.

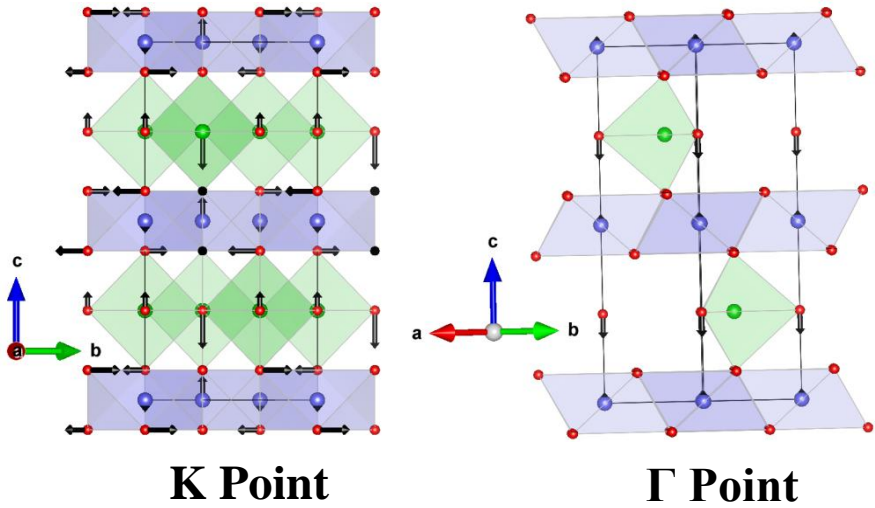

**K Point**   **Γ Point**

FIG. 8 (a) The calculated mode Grüneisen parameters from the pressure dependence of phonon frequencies in low temperature phase (space group P6$_3$cm) of YMnO$_3$, within the local density approximation (LDA). (b) The computed and experimental [11, 59] volume thermal expansion behavior in the low temperature phase (space group P6$_3$cm) of YMnO$_3$, within the local density approximation (LDA).

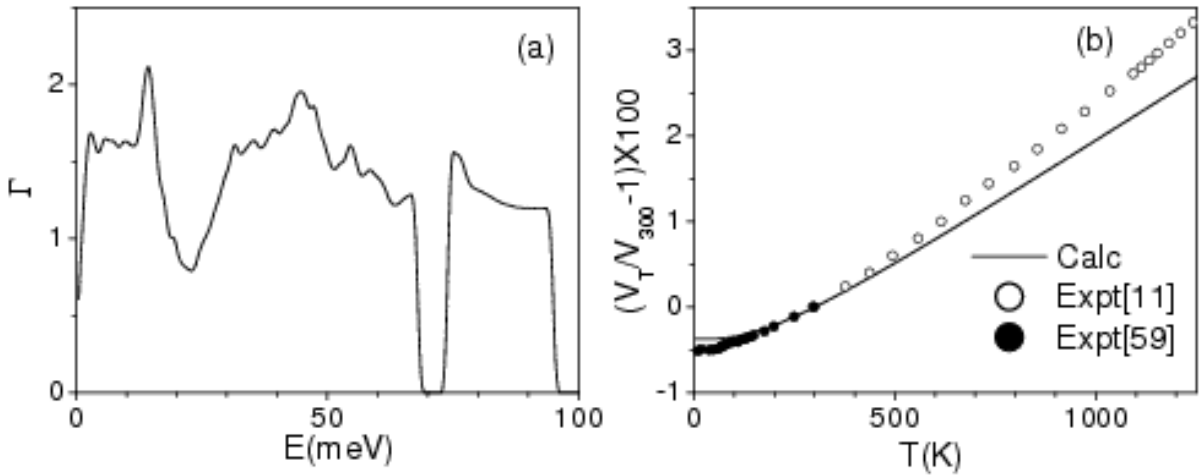